\documentclass[a4paper,preprint]{aastex}

\usepackage{natbib}

\begin{document}

\title{A Search for Additional Planets in Five of the Exoplanetary Systems Studied by the NASA {\it EPOXI} Mission}

\author{Sarah~Ballard\altaffilmark{1}, Jessie~L.~Christiansen\altaffilmark{2}, David~Charbonneau\altaffilmark{1}, Drake~Deming\altaffilmark{3}, Matthew~J.~Holman\altaffilmark{1}, Michael~F.~A'Hearn\altaffilmark{4}, Dennis~D.~Wellnitz\altaffilmark{4}, Richard~K.~Barry\altaffilmark{3}, Marc~J.~Kuchner\altaffilmark{3}, Timothy~A.~Livengood\altaffilmark{3}, Tilak~Hewagama\altaffilmark{3,4}, Jessica~M.~Sunshine\altaffilmark{4}, Don~L.~Hampton\altaffilmark{5}, Carey~M.~Lisse\altaffilmark{6}, Sara~Seager\altaffilmark{7}, and Joseph~F.~Veverka\altaffilmark{8}}

\altaffiltext{1}{Harvard-Smithsonian Center for Astrophysics, 60 Garden Street, Cambridge, MA 02138; sballard@cfa.harvard.edu}
\altaffiltext{2}{NASA Ames Research Center, Moffett Field, CA 94035}
\altaffiltext{3}{NASA/Goddard Space Flight Center, Greenbelt, MD 20771}
\altaffiltext{4}{University of Maryland, College Park, MD 20742}
\altaffiltext{5}{University of Alaska Fairbanks, Fairbanks AK 99775}
\altaffiltext{6}{Johns Hopkins University Applied Physics Laboratory, Laurel, MD 20723}
\altaffiltext{7}{Massachusetts Institute of Technology, Cambridge, MA 02139}
\altaffiltext{8}{Cornell University, Space Sciences Department, Ithaca, NY 14853}

\begin{abstract}
We present time series photometry and constraints on additional planets in five of the exoplanetary systems studied by the EPOCh (Extrasolar Planet Observation and Characterization) component of the NASA {\it EPOXI} mission: HAT-P-4, TrES-3, TrES-2, WASP-3, and HAT-P-7. We conduct a search of the high-precision time series for photometric transits of additional planets. We find no candidate transits with significance higher than our detection limit. From Monte Carlo tests of the time series using putative periods from 0.5 days to 7 days, we demonstrate the sensitivity to detect Neptune-sized companions around TrES-2, sub-Saturn-sized companions in the HAT-P-4, TrES-3, and WASP-3 systems, and Saturn-sized companions around HAT-P-7. We investigate in particular our sensitivity to additional transits in the dynamically favorable 3:2 and 2:1 exterior resonances with the known exoplanets: if we assume coplanar orbits with the known planets, then companions in these resonances with HAT-P-4b, WASP-3b, and HAT-P-7b would be expected to transit, and we can set lower limits on the radii of companions in these systems. In the nearly grazing exoplanetary systems TrES-3 and TrES-2, additional coplanar planets in these resonances are not expected to transit. However, we place lower limits on the radii of companions that would transit if the orbits were misaligned by 2.0$^{\circ}$ and 1.4$^{\circ}$ for TrES-3 and TrES-2, respectively.
\end{abstract}


\section{Introduction}
{\it EPOXI} (EPOCh + DIXI) is a NASA Discovery Program Mission of Opportunity using the Deep Impact flyby spacecraft \citep{Blume05}. From January through August 2008, the EPOCh (Extrasolar Planet Observation and Characterization) Science Investigation used the HRI camera \citep{Hampton05} with a broad visible bandpass filter to gather precise, rapid cadence photometric time series of known transiting exoplanet systems. The majority of these targets were each observed nearly continuously for several weeks at a time. In Table 1 we give basic information about the seven EPOCh targets and the number of transits of each that EPOCh observed. 

 One of the EPOCh science goals is a search for additional planets in these systems. Such planets would be revealed either through the variations they induce on the transits of the known exoplanet, or directly through the transit of the second planet itself. The search for additional planets in the EPOCh observations of the M~dwarf GJ~436 was presented in \cite{Ballard10}. Because GJ~436 is a nearby M~dwarf, it is the only EPOCh target for which we are sensitive to planets as small as 1.25 $R_{\oplus}$. In this work, we conduct a search for photometric transits of additional planets; the transit times of the known exoplanets observed by EPOCh are presented in \cite{Christiansen10b}.

The search for transiting planets in the EPOCh light curves is scientifically compelling because the discovery of two transiting bodies in the same system permits the direct observation of their mutual dynamical interactions. This enables constraints on the masses of the two bodies independent of any radial velocity measurement \citep{Holman05, Agol05}, as has been done for the multiple transiting planet system Kepler 9 \citep{Holman10}. There are also separate motivations for searches for additional planets around the EPOCh targets. The search for additional transits in the EPOCh observations is complementary to existing constraints on additional planets from photometric observations, radial velocity measurements, and transit timing analyses of the known exoplanet.

Here we briefly summarize such work to date for our five targets. \cite{Smith09} investigated 24 light curves of known transiting exoplanets, including the {\it EPOXI} targets HAT-P-4, TrES-2, WASP-3, and HAT-P-7, and found that they were sensitive to additional transits of Saturn-sized planets with orbital periods less than 10 days with greater than 50\% certainty, although that probability is less for HAT-P-4 \citep{Kovacs07} because of decreased phase coverage. Transit timing analyses of TrES-3b \citep{ODonovan07} have ruled out planets in interior and exterior 2:1 resonances \citep{Gibson09}, although the transit times obtained by \cite{Sozzetti09} for TrES-3b may suggest a deviation from a constant period that could be attributed to an additional body. \cite{Freistetter09} found that a broad range of orbits around TrES-2 \citep{ODonovan06} would be dynamically stable for additional planets, although the constraints presented by \cite{Rabus09} for TrES-2 have ruled out an 5 $M_{\oplus}$ planet in the 2:1 resonance specifically, and \cite{Holman07} found no deviations in the transit timing residuals from the predicted ephemeris. Additionally, \cite{Raetz09} observed a candidate transit in their photometry of TrES-2 which might be attributed to an additional body in the system in an external orbit to TrES-2b. However, \cite{Kipping10} investigated the TrES-2 {\slshape Kepler} observations and found no unexpected photometric decrements and no significant transit timing or transit duration variation. \cite{Maciejewski10} performed an analysis of the transit times of WASP-3b \citep{Pollacco08}, and found evidence for planet with a mass of 15 $R_{\oplus}$ in a orbit close to a 2:1 resonance with the known planet. In the HAT-P-7 \citep{Pal08} radial velocity measurements, \cite{Winn09} found a drift that provides evidence for a third body. This radial velocity trend is consistent with any period longer than a few months. Finally, the light curves obtained by the {\slshape Kepler} Mission \citep{Borucki10} will ultimately enable exquisite constraints on the presence of additional planets in two of the systems which were also observed by {\it EPOXI}: TrES-2 and HAT-P-7.

The remainder of this paper is organized as follows. In Section 2, we describe the photometry pipeline we created to produce the time series. In Section 3, we describe the search we conduct for additional transiting planets. We present a Monte Carlo analysis of the EPOCh observations of HAT-P-4, TrES-3, TrES-2, WASP-3, and HAT-P-7, and demonstrate the sensitivity to detect additional planets in the Neptune-sized and Saturn-sized radius ranges. In Section 4, we present our best candidate transit signals, and from the search for additional transits we place upper limits on the radii of additional putative planet in these systems in the exterior 3:2 and 2:1 resonances with the known exoplanets.

\section{Observations and Data Reduction}
 
The photometric pipeline we developed for the EPOCh data is discussed at length in \cite{Ballard10} (concerning GJ~436 in particular), and is summarized in \cite{Christiansen10a} (concerning HAT-P-7) and \cite{Christiansen10b} (concerning HAT-P-4, TrES-3, TrES-2, and WASP-3). We outline here the basic steps we undertake to produce the final EPOCh time series. We acquired observations of the five EPOCh targets presented here nearly continuously for approximately two-week intervals. These intervals were interrupted for several hours at approximately 2-day intervals for data downloads. We also obtained for TrES-2, WASP-3, and HAT-P-7 approximately one day of ``pre-look'' observations, implemented to optimize pointing for each target, that predate the continuous observations by a week.  The basic characteristics of the targets and observations are given in Tables 1 and 2. Observations of this type were not contemplated during development of the original Deep Impact mission; the spacecraft was not designed to maintain very precise pointing over the timescale of a transit (Table \ref{tbl-2}). Furthermore, the available onboard memory precludes storing the requisite number of full-frame images (1024$\times$1024 pixels).  Hence, the observing strategy during the later observations (TrES-2, WASP-3, and HAT-P-7) used 256$\times$256 sub-array mode for those times spanning the transit, and 128$\times$128 otherwise.  This strategy assured complete coverage at transit, with minimal losses due to pointing jitter exceeding the 128$\times$128 sub-array at other times. We elected to exclude the following EPOCh data from the search for additional transits: first, the observations of XO-2, for which we gathered only partial transits and had relatively sparse phase coverage due to pointing jitter and data transfer losses. Second, we did not use the observations from the second EPOCh campaign for HAT-P-4 (from 29 Jun - Jul 7 2008), which we could not calibrate to the same level of precision as the original observations for reasons explained below. Our sensitivity to additional transit signals in the HAT-P-4 light curve, which should theoretically have improved with additional observations removed in time, was in reality diminished due to the increased correlated noise in the second campaign HAT-P-4 observations. For this reason, we elected to use only the original 22 days of observations in the search for additional transits. 

\begin{deluxetable}{cccc}
\tabletypesize{\scriptsize}
\tablecaption{EPOCh Targets
\label{tbl-1}}
\tablewidth{0pt}
\tablehead{
\colhead{Name} & \colhead{$V$ Magnitude} & 
\colhead{Number of Transits Observed\tablenotemark{a}} & \colhead{Dates Observed [2008]}
}
\startdata
HAT-P-4 & 11.22 & 10 & Jan 22--Feb 12, Jun 29--Jul 7 \\ 
TrES-3 & 11.18 & 7 & Mar 8--March 10, March 12--Mar 18\\
XO-2 & 12.40 & 3  & Mar 11, Mar 23--Mar 28\\ 
GJ 436 & 10.67 & 8 & May 5--May 29\\
TrES-2 & 11.41 & 9 & Jun 27--Jun 28, Jul 9--Jul 18, Jul 21--Aug 1\\
WASP-3 & 10.64 & 8 & Jul 18--Jul 19, Aug 1--Aug 9, Aug 11--Aug 17 \\
HAT-P-7 & 10.50 & 8 & Aug 9--Aug 10, Aug 18--Aug 31\\
\enddata
\tablenotetext{a}{Some transits are partial.}
\end{deluxetable}

\begin{deluxetable}{cc}
\tabletypesize{\scriptsize}
\tablecaption{Characteristics of the EPOCh Observations 
\label{tbl-2}}
\tablewidth{0pt}
\tablehead{
\colhead{} & \colhead{}  
}
\startdata
Telescope aperture & 30 cm \\ 
Spacecraft memory & 300 Mb \\
Bandpass & 350-1000 nm \\
Integration time & 50 seconds \\
Pointing jitter  & $\pm$ 20 arc-sec per hour \\ 
Defocus & 4 arc-sec FWHM \\
Pixel scale & 0.4 arc-sec per pixel \\ 
Subarray size & 256$\times$256 pixels spanning transit, 128$\times$128 otherwise\tablenotemark{a} \\
\enddata
\tablenotetext{a}{With the exception of the HAT-P-4 observations during 2008 January and February and TrES-3 observations, which were conducted entirely in 128$\times$128 subarray mode.}
\end{deluxetable}

We used the existing Deep Impact data reduction pipeline to perform bias and dark subtractions, as well as preliminary flat fielding \citep{Klaasen05}. We first determined the position of the star on the CCD using PSF fitting, by maximizing the goodness-of-fit (with the $\chi^{2}$ statistic as an estimator) between an image and a model PSF (oversampled by a factor of 100) with variable position, additive sky background, and multiplicative brightness scale factor. We then processed the images to remove sources of systematic error due to the CCD readout electronics. We first scaled down the two central rows by a constant value, then we scaled down the central columns by a separate constant value. Finally, in the case of 256$\times$256 images, we scaled the entire image by a multiplicative factor to match the 128$\times$128 images (the determination of the optimal correction values is performed independently for each target). We performed aperture photometry on the corrected images, using an aperture radius of 10 pixels, corresponding to twice the HWHM of the PSF. To remove remaining correlated noise due to the interpixel sensitivity variations on the CCD, we fit a 2D spline surface to the brightness variations on the array as follows. We randomly drew a subset of several thousand out-of-transit and out-of-eclipse points from the light curve (from a data set ranging from 11,000 total points in the case of TrES-3 to 20,000 points in the case of HAT-P-4), recorded their X and Y positions, and calculated a robust mean of the brightness of the 30 nearest spatial neighbors for each selected point. To determine the robust mean, we used an iterative sigma-clipping routine that recalculates the mean after excluding outliers further than 3 sigma from the mean estimate at each iteration (the routine concludes after the iteration when no new outliers are identified). Given the set of X and Y positions and the average brightness values of the 30 points which lie nearest those positions, we fit a spline surface to the brightness variations in X and Y using the IDL routine \verb=GRID_TPS=. This spline surface has the same resolution as the CCD, and approximates a flat field of the CCD which has been convolved by a smoothing kernel with width equal to the average distance required to enclose 30 neighboring points. We then corrected each data point individually by bilinearly interpolating on the spline surface to find the expected brightness of the star at each X and Y position. We then divide each observation by its expected brightness to remove the effects of interpixel sensitivity variations. We used only a small fraction of the observations to create the spline surface in order to minimize the potential transit signal suppression introduced by flat fielding the data with itself. To produce the final time series, we iterated the above steps, fitting for the row and column multiplicative factors, the sub-array size scaling factor, and the 2D spline surface that minimized the out-of-transit standard deviation of the photometric time series.

We include two additional steps in the reduction of these data that were not included in the \cite{Ballard10} reduction of the GJ~436 EPOCh observations. First, during the second campaign observations of HAT-P-4 and TrES-2, we observed an increase in brightness when the position of the star was located in the lower right-hand quadrant of the CCD. At the image level, we observed a bright striping pattern in this quadrant that caused the measured brightness of the star to increase as soon as the PSF entered this quadrant. We found that the dependence of the brightness increase in this quadrant was correlated with the Instrument Control Board Temperature value recorded for for each image in the FITS header. For the HAT-P-4 second campaign and TrES-2 observations, we first fit a spline to the dependence of the photometric residuals (after the bootstrap flat field was applied, described below) on the Instrument Control Board Temperature, using residuals for which the entire PSF of the star fell into the CCD quadrant in question. The most egregious brightness increase is 4 mmag, when this effect was most prominent. We then performed aperture photometry again on these targets and corrected each image by interpolating the Instrument Control Board Temperature for that image onto the spline, multiplying this correction value by the fraction of the PSF core that fell into the quadrant in question, and dividing this value from the photometry. We found that this iterative procedure largely removed this quadrant-dependent effect. In the latter half of TrES-2 observations, we no longer observed the brightness increase in this CCD quadrant. Therefore, we found that the correction procedure was only necessarily for the second campaign HAT-P-4 and the first portion of TrES-2 observations. However, because of the 6-month separation of the second campaign HAT-P-4 observations from the original HAT-P-4 observations, the behavior of the CCD had sufficiently altered to disallow the combination of the data into a single 2D spline surface. The separate 2D spline correction of the second campaign observations (spanning only 8 days), coupled with the residual striping artifacts, sufficiently decreased the precision of the second campaign observations that we elected to exclude them from the search for additional transiting planets around HAT-P-4.

Secondly, we include one final correction after we have removed the interpixel brightness variations with the 2D spline, which is to perform an additional point-by-point correction to the data taken during transit and secondary eclipse of the known exoplanets. The bootstrap flat field randomly selects a set of points to create the spline surface, instead of using all the data to create this surface; this minimizes the suppression of additional transits. Our sensitivity to additional transits is sufficiently diminished during the transits of the known planet that we are concerned more with removing correlated noise, and less concerned about avoiding additional transit suppression. The two reasons for the diminished sensitivity during transit windows are, first, that we fit a slope with time to the points immediately outside of the transit of the known exoplanet (from 3 minutes to 30 minutes before and after transit) and divide by the slope in order to normalize each transit before we fit for the system parameters (this procedure is also detailed in \citealt{Ballard10}). This could have the possibility of removing a decrement due to an additional transit. Second, there is also the possibility of an occultation of one planet by another. We therefore elected to perform a point-by-point correction after the 2D spline for points occurring during the transit and eclipse of the known exoplanet, wherein we find a robust mean of the 30 nearest neighbors to each point (using the same iterative routine described above) and divide this value individually for each point in transit or eclipse. This has the benefit of removing additional correlated noise during transit and eclipse, while still minimizing signal suppression of additional putative planets outside of these time windows

\begin{figure}[h!]
\begin{center}
 \includegraphics[height=6.9in]{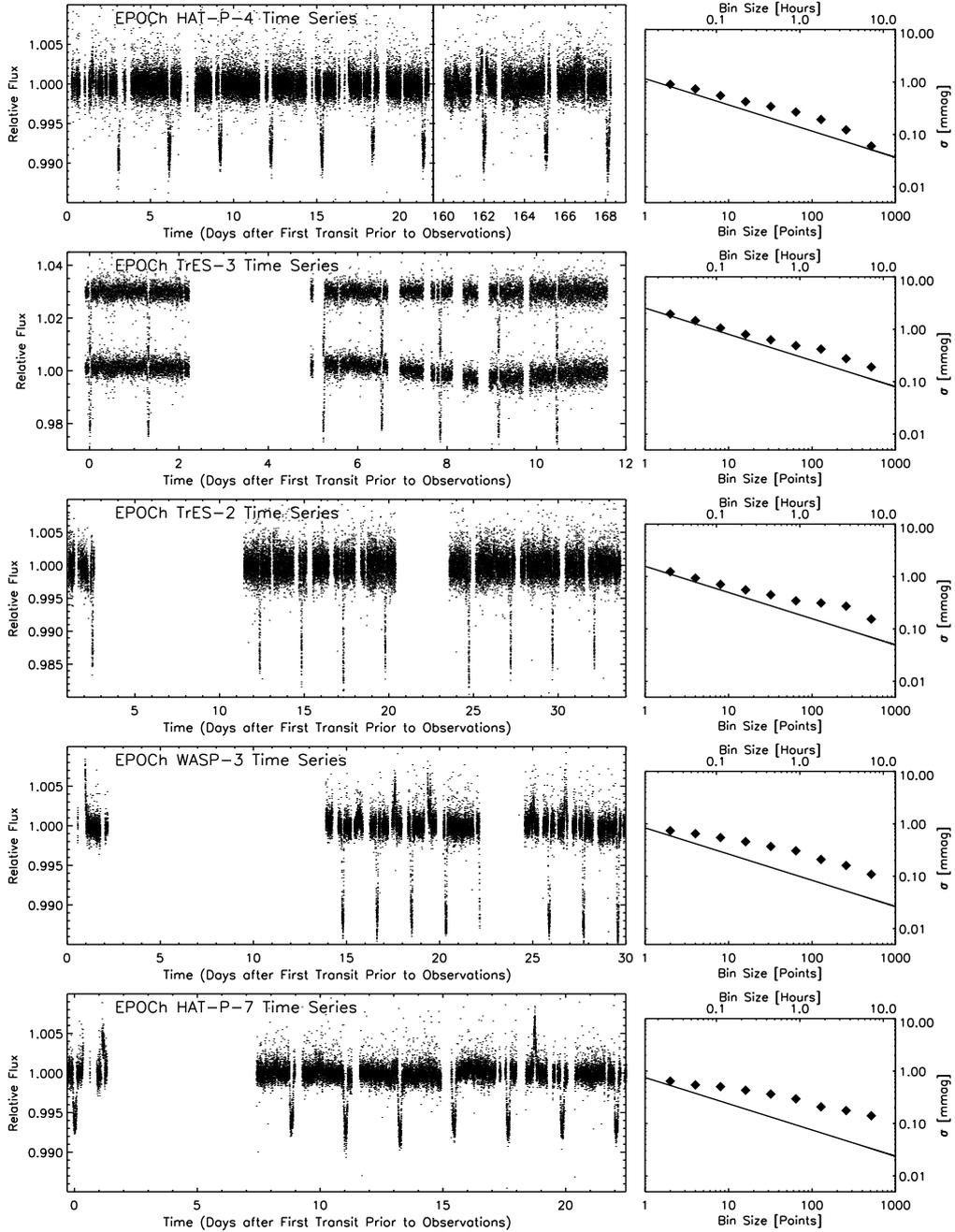} 
 \caption{\textit{Left panels:} {\it EPOXI} time series for targets HAT-P-4, TrES-3, TrES-2, WASP-3, and HAT-P-7. For TrES-3 (second panel from top), we show the light curve with original modulation due to star spots at bottom. \textit{Right panels:} The standard deviation versus bin size for each target, compared to the ideal Gaussian limit (shown with a line, normalized to match the value at N=1).}
   \label{fig:lightcurve1}
\end{center}
\end{figure}

After we take these steps to address the systematics associated with the observations, we achieve a precision for the unbinned observations which is approximately twice the photon noise limited precision for all five targets. We estimate the photon limited precision at the image level, by converting the stellar flux from watts per meter squared per steradian per micron (W/m$^{2}\cdot$sr$\cdot\mu$m) to electron counts as follows. We first divide by the conversion factor in the FITS header (keyword RADCALV = 0.0001175 W/m$^{2}\cdot$sr$\cdot\mu$m per DN/s), then multiply by the exposure time (INTTIME = 50.0005 s), and finally multiply by the gain (28.80 e/DN, per \citealt{Klaasen08} for the HRI camera). We then estimate the photometric error by calculating $1/\sqrt{N}$, where N is the number of electrons. We have excluded read noise, bias, and dark current from the estimation of the photon limited precision because these quantities contribute negligibly to the total number of electrons measured within the aperture; we briefly summarize our reasoning here. Using the results of the calibration tests on the HRI instrument shown in \cite{Klaasen08}, we estimate the read noise and dark current (given the CCD temperature of 160 K, as recorded in the image headers) to contribute less than a DN. Calculating the median bias value per pixel from the overclocked pixels associated with each image, and then multiplying this median value by the number of pixels contained within the aperture, we determine that the bias contributes less than 500 DN. When compared to the total measured DN flux contained in the aperture, which is of order 10$^{5}$ DN for the dimmest target star, TrES-3, we conclude that read noise, bias, and dark current are negligible. We repeat the photon limited precision calculation on 50 images for each target, and take the mean of these values to be our estimate for the photon limited precision. Our precision of 1.21 mmag for HAT-P-4 is 94\% above the limit, 2.17 mmag for TrES-3 is 106\% above the limit, 1.62 mmag for TrES-2 is 136\% above the limit, 0.97 mmag for WASP-3 is 106\% above the limit, and 0.86 mmag for HAT-P-7 is 91\% above the limit.  The EPOCh precision for GJ~436 of 0.51 mmag was only 56\% above the photon noise limit, which we attribute to the longer baseline of observations with fewer gaps in phase coverage, both of which enabled us to create a higher quality 2D spline flat field \citep{Ballard10}. Figure \ref{fig:lightcurve1} shows five EPOCh time series after the 2D spline correction is applied; these light curves are identical to the ones presented in \cite{Christiansen10a,Christiansen10b}. In the right panel adjacent to each time series, we show how the time series, after the 2D spline is applied, bins down as compared to Gaussian noise over timescales of 7 hours (512 points) or less. We selected the longest contiguous portion of the lightcurve between transits (and excluding secondary eclipse) to calculate the standard deviation as a function of binsize-- this unbinned portion typically comprises about 2500 points. We compare the expected Gaussian scatter for a bin size of 1 hour (assuming that sigma decreases as ${N}^{-1/2}$, and normalizing to the observed rms of the unbinned data) to the measured scatter, and find that the presence of correlated noise inflates the 1$\sigma$ error bar by a factor of 1.86 for HAT-P-4, 1.58 for TrES-3, 1.90 for TrES-2, 2.77 for WASP-3, and 3.14 for HAT-P-7 for 1 hour timescales.

\begin{figure}[h!]
\begin{center}
 \includegraphics[height=6.5in]{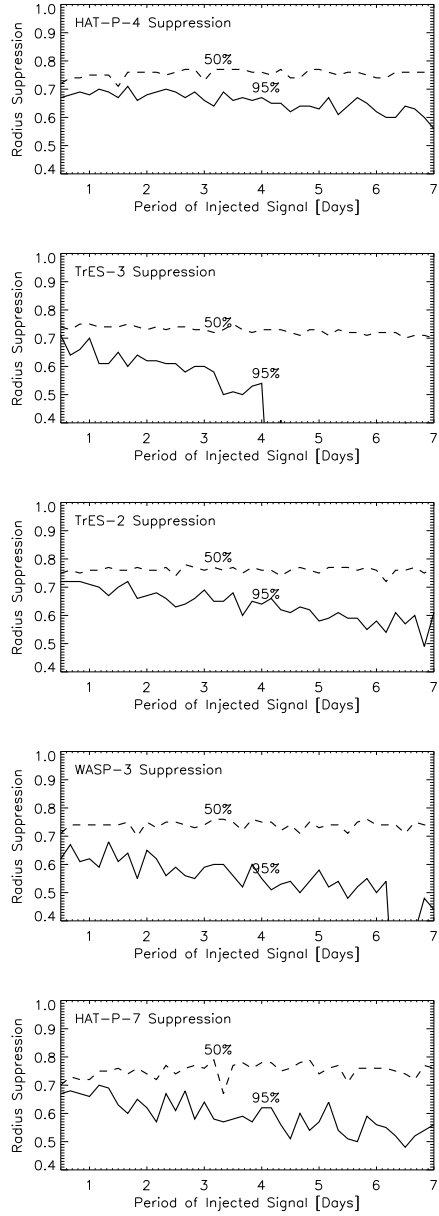} 
 \caption{The 50\% and 95\% confidence values for suppression of additional transits as a function of orbital period in the EPOCh observations. We have 50\% confidence that the transit signal will not be suppressed more than the value of the dashed line at that period, and  95\% confidence that the transit signal will not be suppressed more than the value of the solid line.}
   \label{fig:suppression}
\end{center}
\end{figure}

We also investigate the transit signal suppression introduced by using a flat field created from the out-of-transit and out-of-eclipse data itself. We avoid the suppression of known transits in each data set by iteratively excluding those observations (using an ephemeris for the known planet derived from the EPOCh observations) from the points used to generate the flat field surface, so that we only use the presumably constant out-of-transit and out-of-eclipse observations to sample the CCD sensitivity. However, if the transit of an additional planet occurs while the stellar PSF is lying on a part of the CCD that is never visited again, the 2D spline algorithm instead models the transit as diminished pixel sensitivity in that CCD location. To quantify the suppression of additional transits that result from using the 2D spline, we inject transit light curves with periods ranging from 0.5 days to 7 days in intervals of 30 minutes in phase (ranging from a phase of zero to a phase equal to the period) into the EPOCh light curve just prior to the 2D spline step. After performing the 2D spline, we then phase the data at the known injected period and fit for the best radius, using $\chi^{2}$ as the goodness-of-fit statistic. We show in Figure \ref{fig:suppression} the radius suppression as a function of period for five EPOCh targets.  The HAT-P-4 observations occur over a longer duration with less gaps in phase coverage, so even at a period of 7 days, we have 95\% confidence that the radius of an additional transiting planet will not be suppressed to less than 60\% its original value. For example, an additional 8 $R_{\oplus}$ planet orbiting HAT-P-4 will appear no smaller than 0.6$\times$8 $R_{\oplus}$, or 4.8 $R_{\oplus}$, with 95\% confidence. However, for a target with sparser phase coverage, such as WASP-3, we have 95\% confidence that the radius will not be suppressed to less than 45\% its original value. The same 8 $R_{\oplus}$ planet orbiting WASP-3 will therefore appear no smaller than 0.45$\times$8 $R_{\oplus}$, or 3.6 $R_{\oplus}$, with the same confidence. The average (50\% confidence) suppression value of 75\% across all periods and for all targets reflects the average density of points on the CCD (and thus the quality of the 2D spline), which is indicative of the pointing jitter of the instrument.  We describe our incorporation of signal suppression into our search for additional planets in greater detail in Section 3.2.

\section{Analysis}

\subsection{Search for Additional Transiting Planets}
We search the {\it EPOXI} time series for evidence for additional shallow transits. We developed software to search for these additional transits, which is discussed at length in \cite{Ballard10}. The steps involved in the procedure are summarized in this section. We conduct a Monte Carlo analysis to assess how accurately we could recover an injected planetary signal in each of the EPOCh light curves. We evaluate our sensitivity to transit signals on a grid in radius and period space sampled at regular intervals in $R_{P}^{2}$ and regular frequency spacing in $P$. We create an optimally spaced grid as follows: for the lowest period at each radius, we determine the radii at which to evaluate the adjacent periods by solving for the radius at which we achieve equivalent signal-to-noise (for this reason, we expect significance contours to roughly coincide with the grid spacing). We use the \cite{Mandel02} routines for generating limb-darkened light curves given these parameters to compute a grid of models corresponding to additional possible planets. If we make the simplifying assumptions of negligible limb darkening of the host star, a circular orbit, and an orbital inclination angle $i$ of 90$^{\circ}$, the set of light curves for additional transiting bodies is a three parameter family. These parameters are radius of the planet $R_{p}$, orbital period of the planet $P$, and orbital phase $\phi$. To generate these light curves, we also use the stellar radii values determined by \cite{Christiansen10a,Christiansen10b} from the {\it EPOXI} data, with the corresponding stellar masses, from the literature, that were used to calculate those radii. Those radius and mass values are 1.60 $R_{\odot}$ and 1.26 $M_{\odot}$ \citep{Kovacs07} for HAT-P-4, 0.82 $R_{\odot}$ and 0.93 $M_{\odot}$ \citep{Sozzetti09} for TrES-3, 0.94 $R_{\odot}$ and 0.98 $M_{\odot}$ \citep{Sozzetti07} for TrES-2, 1.35 $R_{\odot}$ and 1.24 $M_{\odot}$ \citep{Pollacco08} for WASP-3, and 1.82 $R_{\odot}$ and 1.47 $M_{\odot}$ \citep{Pal08} for HAT-P-7.  At each test radius and period, we inject planetary signals with 75 randomly assigned phases into the residuals of EPOCh light curves with the best transit model divided out, and then attempt to recover blindly the injected signal by minimizing the $\chi^{2}$ statistic. The period range of injected signals is selected for each target individually, to ensure the injected transit signal comprises at least two transits in most cases. For a target with high phase coverage, like HAT-P-4, we inject signals with periods up to 7 days, but for targets with observations of a shorter duration and sparser phase coverage, like TrES-3 or WASP-3, we inject signals up to 3.5 and 2.5 days, respectively. For planets with periods higher than these values, we may detect the planet, but with a single transit, we expect only a very weak constraint on the period.

We first conduct a coarse $\chi^{2}$ grid search in radius, period, and phase. We select the spacing of this grid to minimize processing time while ensuring that the transit was not missed; we polish the parameters with a finer $\chi^{2}$ search after the initial coarse search. We sample the $\chi^{2}$ space at 300 points in period space (at even frequency intervals between 0.5 and 8.5 days), 50 points in radius space (between 0.5 and 5.5 Earth radii) and a variable number of points in phase space set by the resolution of the transit duration for each period. We use an expression for the transit duration $\tau$ given by \cite{Seager03}:

\begin{equation}
\mbox{sin }i\mbox{ sin}\left(\frac{\pi\tau}{P}\right)=\sqrt{\left( \frac{R_{\star}+R_{P}}{a}\right)^{2}-\mbox{cos}^{2}i}.
 \label{eq:duration}
\end{equation}

For each test model, we compute the $\chi^{2}$, using the out-of-transit standard deviation to estimate the error in each point. After the grid $\chi^{2}$ minimum is determined, we use the \verb=amoeba= minimization routine \citep{Nelder65} to more finely sample the $\chi^{2}$ space in order to find the $\chi^{2}$ minimum from the specified nearest grid point. We also investigate whether aliases of the best-fit period from the $\chi^{2}$ grid improve the fit. We find that roughly half of the best solutions from the grid are aliases of the injected period, most at either half or twice the value of the injected period, but we test aliases at every integer ratio from 1/35 to 35 times the given period (although aliases other than 1:2, 2:1, 3:1, 1:3, 2:3, or 3:2 occurs less than 3\% of the time for all targets). We also repeat the finer grid search at the three next lowest $\chi^{2}$ minima, in case the best solution (or an alias of the best solution) lies closer to that grid point. For all injected signals, we recover a solution which is a better fit (in the $\chi^{2}$ sense) than the injected signal. For this reason, we are confident that we are sampling the $\chi^{2}$ space sufficiently finely to locate the best solution. 

\begin{figure}[h!]
\begin{center}
 \includegraphics[height=6.5in]{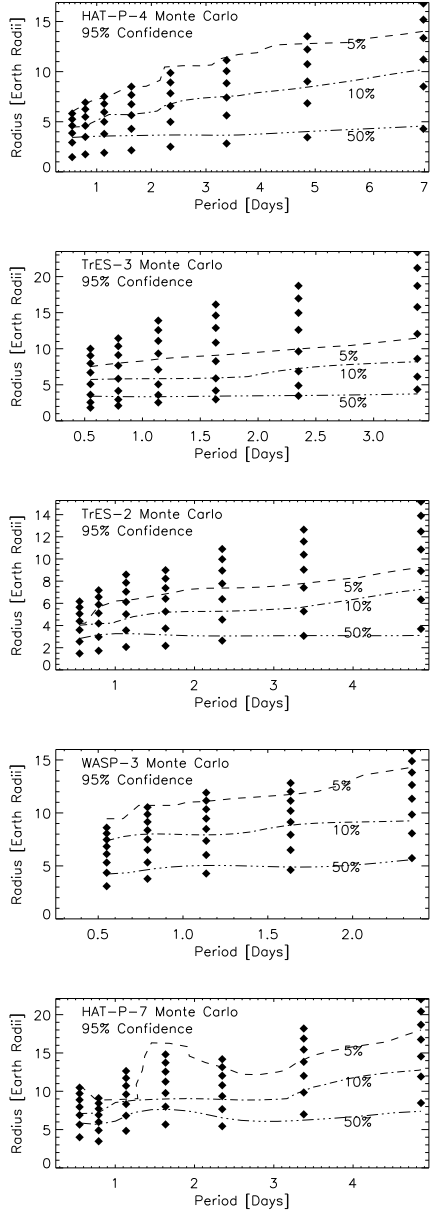} 
 \caption{Constraints on radius from the Monte Carlo analysis. For each point in radius and period, we create 75 light curves with random orbital phases, inject them into the EPOCh residuals, and attempt to recover them blindly. The diamonds indicate the grid of radii and periods at which we evaluate our sensitivity; the contours are produced by interpolating between these points. The contours indicate the relative error in radius (absolute value of recovered-injected/injected radius) that encloses 95\% of the results.}
  \label{fig:success_results_all}
\end{center}
\end{figure}

We quantify the success of this analysis by how well the search blindly recovers the known injected transit signal. We define the error on the recovered parameter, for instance period, to be $\mid P_{injected}-P_{observed}\mid/P_{injected}$. Figure \ref{fig:success_results_all} shows this relative error in radius, with 95\% confidence, for all searches.  As we note in the last paragraph of Section 2, we anticipate suppression of additional transit signals from the bootstrap flat field treatment of the {\it EPOXI} data. We evaluate the suppression we expect at the period values used in the Monte Carlo analysis, using the results shown in Figure \ref{fig:suppression}. We incorporate this expected suppression by vertically shifting the effective radius values of the grid points at which we evaluate our sensitivity to additional transits. For example, for HAT-P-4 at 1.63 days, all grid points have been shifted upward in radius by a value of 1/0.7, or 1.42, because we anticipate that the radius will be suppressed to no smaller than 70\% its original value. For this reason, the recovery statistics corresponding to a 3.0 $R_{\oplus}$ transit depth in the final light curve would be accurate for an original transit signal of a 4.3 $R_{\oplus}$ planet once we fold our expectation of signal suppression. 

\begin{figure}[h!]
\begin{center}
 \includegraphics[height=6.5in]{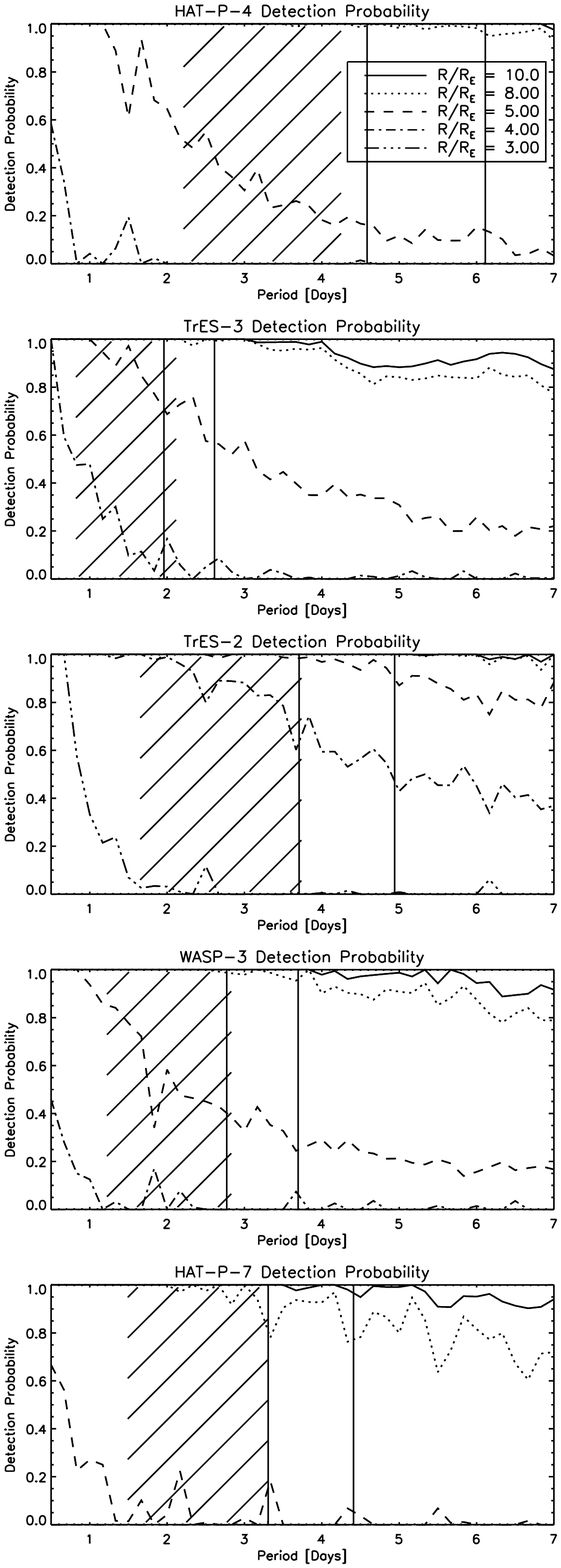} 
 \caption{Detection probability versus period for planets ranging in size from 3 to 10 $R_{\oplus}$. The detection criteria is set by the percentage of phases at a given period for which the number of points observed in transit produces a $\chi^{2}$ improvement of the cutoff significance, compared to the null hypothesis ($\Delta\chi^{2}$ of 250 for HAT-P-4, 200 for TrES-3 and TrES-2, 400 for WASP-3, and 500 for HAT-P-7). We assume a boxcar-shaped transit at the depth of $(R_{P}/R_{\star})^{2}$. The vertical lines show the positions of the 3:2 and 2:1 resonances with the known planet, and the cross-hatching shows the location of orbits which are not guaranteed to be stable by Hill's criterion per \cite{1993G}.}
   \label{fig:coverage}
\end{center}
\end{figure}

We also evaluate the overall detection probability for putative transiting planets. Given the cadence and coverage of the {\it EPOXI} observations, we determine the number of in-transit points we expect for a given radius, period, and phase (where the phase is evaluated from 0 to 1 periods, in increments of 30 minutes). We then evaluate the expected significance of the detection, assuming a boxcar-shaped transit at the depth of $(R_{P}/R_{\star})^{2}$, and the standard deviation of the time series. At each phase and period individually, we scale down $R_{P}$ to incorporate the signal suppression at that ephemeris. We use the improvement in the $\chi^{2}$ over the null hypothesis to define a positive detection, after we have removed the best candidate transit signal (described in Section 4.1). If we do not first remove this signal, then we are a priori defining a ``detectable'' signal to be any signal more prominent than the best candidate signal, and we would be unable to evaluate this signal's authenticity. We set our detection limit at an improvement in $\chi^{2}$ over the null hypothesis that signifies a correctly recovered period (which we define as a period error of $<$1\%). This number is variable among the EPOCh targets due to the precision of the observations and the contamination of correlated noise. The detection probability of additional transiting planets, as a function of their radius and orbital period, is shown in Figure \ref{fig:coverage}.

For the HAT-P-4, TrES-3, and TrES-2, the $\Delta\chi^{2}$ cutoff is set at 250, 200, and 200 respectively. For WASP-3, the $\Delta\chi^{2}$ criterion for detection is 400, and for HAT-P-7, the cutoff is 500. There are five exceptions here for these threshold values across the five targets: in our analysis of WASP-3, we find two instances of a significance higher than 400, but an incorrect period value: these signals both comprise 4 full transit events, and are recovered at a 4:1 alias of the true period of 2.34 days. Due to the same instances of correlated noise that produce two positive deviations during two of the four transit events that decrease the depths by 2 mmag, a better solution is found at an alias of 4:1 that at the true period. For HAT-P-7, we find three similar cases of a 2:1 alias providing a better solution than the injected period, although the significance of the detection is above the threshold vale of 500. For these three injected signals, which comprise three transits, two of the transits are recovered correctly, and the third overlays a single instance of correlated noise that decreases the depth of the transit by 0.5 mmag. 

We investigated the true signal-to-noise ratio (including correlated noise) associated with a single detectable transit with the cutoff $\Delta\chi^{2}$ significance for each target. We use a method similar to the one described by \cite{Winn08} to determine the contribution of correlated noise to the standard deviation over a transit duration timescale. We first solved for the transit depth associated with the cutoff $\Delta\chi^{2}$ value, assuming a single boxcar transit with standard deviation equal to the out-of-transit and out-of-eclipse standard deviation of the unbinned time series. We next found the standard deviation at a bin size corresponding to a transit duration for each target. We assume an edge-on transit (which assumption we also used for the Monte Carlo analysis) and the shortest period where we expect mostly single transits (this period is slightly larger than the largest period used for the Monte Carlo analysis; which period range was selected so that we would expect at least two transits in nearly all cases). This approximate orbital period is 7.5 days for HAT-P-4, 4.0 days for TrES-3, 5.0 days for TrES-2, 3.0 days for WASP-3, and 5.0 days for HAT-P-7. Using the cutoff transit depth and the standard deviation associated with the transit duration for each target, we find that the signal-to-noise ratio associated with the detection criteria is approximately constant across the targets, ranging between 5 and 8. The variation in the $\Delta\chi^{2}$ value can be attributed in part to the varying presence of correlated noise in the different data sets (and also to the number of points associated with each transit, which depends on the transit duration). We confirm empirically that planets of these radii are detectable by examining the detection probability as a function of radius and orbital period shown in Figure \ref{fig:coverage}. We convert the cutoff transit depth to a planetary radius, assuming the stellar radius derived from the EPOCh observations and average suppression of the radius to 0.75 its original value (roughly constant for all EPOCh targets, as shown in Figure \ref{fig:suppression}). This radius value physically corresponds to the minimum planetary radius detectable by EPOCh from a single transit. This value is 7.1 $R_{\oplus}$ for HAT-P-4, 6.2 $R_{\oplus}$ for TrES-3, 5.3 $R_{\oplus}$ for TrES-2, 6.5 $R_{\oplus}$ for WASP-3, and 7.9 $R_{\oplus}$ for HAT-P-7. Comparing to the nearest radius value in Figure \ref{fig:coverage}, we find that indeed, at the shortest orbital period where we should expect to see single transits, we can detect a planet with radius associated with the detection criteria at high significance. At longer orbit periods, we still expect single transits, but the likelihood that the single transit occurs during a gap in the phase coverage increases.

\section{Discussion}

\subsection{Best Candidate Transit Signals}

\begin{figure}[h!]
\begin{center}
 \includegraphics[height=6.5in]{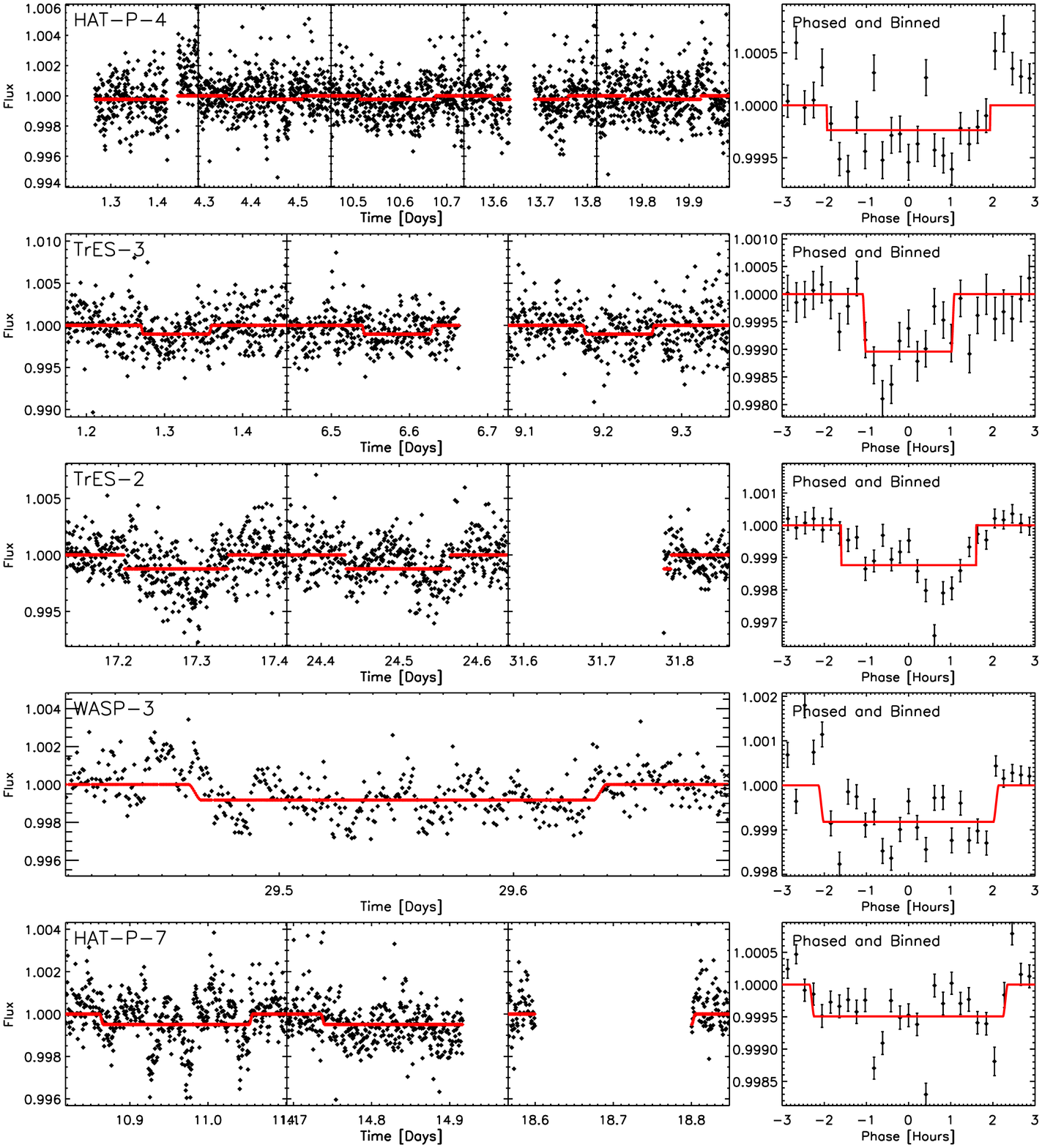} 
 \caption{The best candidate transits for the five EPOCh targets.  Each of the individual candidate transit events comprising the signal are shown at left; the phased and binned signal is shown at right. A time of zero on each X axis corresponds to the time of the first transit of the known planet observed by EPOCh. The $\Delta\chi^{2}$ significance of the HAT-P-4, TrES-3, WASP-3, and HAT-P-7 candidate signals fall below the detection criteria. While the significance of the TrES-2 candidate is above the detection criteria, one of the candidate transits (shown in the leftmost panel) occurs in a sparsely sampled part of the CCD, so the observations are uncalibrated and unreliable. Excising this candidate event, the significance of the remaining signal falls below the detection threshold.}
   \label{fig:best_sols}
\end{center}
\end{figure}

 We present our best candidate transits here, for each of the five EPOCh targets. Figure \ref{fig:best_sols} shows each of the individual candidate transit events that comprise the best candidate signal, as well as the entire phased and binned signal. For HAT-P-4, the best candidate is a 2.7 $R_{\oplus}$ planet in a 3.1 day orbit; the $\Delta\chi^{2}$  significance is 61 (as compared to a detection criterion of 250). For TrES-3, the best candidate is a 2.9 $R_{\oplus}$ planet in a 2.63 day orbit; the $\Delta\chi^{2}$ significance is 87 (as compared to a detection criterion of 200).  For TrES-2, the best candidate is a 3.6 $R_{\oplus}$ planet with a period of 7.22 days; the significance is $\Delta\chi^{2}$ of 269 (as compared to a detection criterion of 200). For WASP-3, the best candidate is a 4.2 $R_{\oplus}$ with a period of 5.9 days; the $\Delta\chi^{2}$ significance is 232 (as compared to a detection criterion of 400). For HAT-P-7, the best candidate is a 4.4 $R_{\oplus}$ planet with a 3.9 day orbit; the significance of the detection is a $\Delta\chi^{2}$ of 201 (as compared to a detection criterion of 500). The only candidate signal above the $\Delta\chi^{2}$ detection threshold is the one in the TrES-2 light curve; this candidate signal comprises two transit events (the other predicted events occur during gaps in the phase coverage). One of the candidate transit events occurs in a portion in the CCD that is never visited afterward; these data are therefore uncalibrated by the 2D spline flat field and are unreliable. Without the transit signal that occurs in the uncalibrated area of the CCD, the $\Delta\chi^{2}$ significance of the remaining transit is 80, which is well below the detection threshold of 200.

\subsection{Radius constraints}

From the results of our Monte Carlo analysis and phase coverage analysis, we can rule out transiting planets in the sub-Saturn radius range for HAT-P-4, TrES-3, and WASP-3, the Saturn-sized radius range for HAT-P-7, and Neptune-sized radius range for TrES-2. We consider in particular our sensitivity to additional planets in the dynamically favorable 3:2 and 2:1 resonance orbits with the known exoplanets. In Figure \ref{fig:coverage}, we show the detection probability as a function of period for planets ranging in size from 3 to 10 $R_{\oplus}$, with positions of  the exterior 3:2 and 2:1 resonances marked by vertical lines. We also indicate in Figure \ref{fig:coverage} the regions not guaranteed to be stable by Hill's criterion, per the formula given in \cite{1993G}. Assuming an eccentricity of zero for both the known and putative additional planet, the planetary orbits are assumed to be stable if the following condition holds, where $\mu_{1}=m_{1}/M_{star}$, $\mu_{2}=m_{2}/M_{star}$, $\alpha=\mu_{1}+\mu_{2}$, and $\delta=\sqrt{a_{2}/a_{1}}$:

\begin{equation}
\alpha^{-3}\left(\mu_{1}+\frac{\mu_{2}}{\delta^{2}}\right)\left(\mu_{1}+\mu_{2}\delta\right)^{2}>1+3^{4/3}\cdot\frac{\mu_{1}\mu_{2}}{\alpha^{4/3}},
 \label{eq:hillradius}
\end{equation}

We solve numerically for the boundaries in $\delta$ of the stable region, using the stellar masses and masses for the known planets given by \cite{Kovacs07} for HAT-P-4b, \cite{Sozzetti09} for TrES-3b, \cite{Sozzetti07} for TrES-2b, \cite{Pollacco08} for WASP-3, and \cite{Pal08} for HAT-P-7, and conservatively using a putative mass for the second body equal to the mass of Saturn. This results in an overestimate of the extent of the Hill-unstable region for the planets with masses smaller than Saturn; while we find we are sensitive to planets with radii well below that of Saturn, the mass of putative additional planets depends on their composition and is uncertain. However, the critical $\delta$ values vary slowly with increased mass of the putative additional planet, so that increasing the mass to that of Jupiter changes the periods associated with the closest Hill-stable orbits by only 7\% at most for these systems. In some cases, the 3:2 orbital resonance is not guaranteed to be Hill-stable (though it may be stable); the exact boundary of the stable region depends on the mass we assume for the additional planet.

From the detection probabilities shown in Figure \ref{fig:coverage}, in the HAT-P-4 system, we are sensitive to planets as small as 8 $R_{\oplus}$ in the 3:2 and 2:1 resonance with HAT-P-4b (with a period of 3.06 days) with 95\% confidence. In the TrES-3 system, with the known exoplanet in a 1.3 day orbit, we would have detected a 5 $R_{\oplus}$ planet in the 3:2 and 1:2 resonance with 70\% and 50\% probability, respectively, and an 8 $R_{\oplus}$ in either orbit with nearly 100\% probability. Around TrES-2 we are sensitive to the smallest planets, and would have detected a 4 $R_{\oplus}$ planet with 65\% probability in the 3:2 resonance with TrES-2b (which has a period of 2.47 days). In both the 3:2 and 2:1 resonances, we had a high probability of detecting a 5.0 $R_{\oplus}$ planet: $>$95\% in the case of the 3:2 resonance, and 90\% in the 2:1 resonance. Around WASP-3, we had 50\% chance of detecting a 5.0 $R_{\oplus}$ planet in the 3:2 resonance with WASP-3b (which has a period of 1.85 days), and would have seen a planet as small as 8 $R_{\oplus}$ in either the 3:2 or 2:1 resonance with $>$95\% probability. Around HAT-P-7, we would have detected a Saturn-sized 10 $R_{\oplus}$ planet in either the 3:2 or 2:1 resonances with 95\% probability, and had a 70\% chance of detecting an 8 $R_{\oplus}$ planet. If we assume an inclination equal to that of the known exoplanet, we can rule out additional transiting planets of HAT-P-4, WASP-3, and HAT-P-7 in the 3:2 and 2:1 resonances of the sizes stated above, as we still expect additional planets to transit at those orbital distances. However, the known exoplanets in both the TrES-3 and TrES-2 systems are already in grazing orbits, so additional planets in the exterior 3:2 and 2:1 resonances would not be expected to transit if they were strictly coplanar with the known exoplanet. However, if the orbits of additional planets were misaligned by 2.0$^{\circ}$ in the case of TrES-3 and 1.4$^{\circ}$ in the case of TrES-2 (using the planetary inclinations and stellar radii from \citealt{Christiansen10a,Christiansen10b}) then we would observe a transit in both of the 3:2 and 2:1 exterior resonances. The orbital inclinations of the ice and gas giants in our solar system vary by up to nearly 2$^{\circ}$ \citep{Cox00}, so it is feasible that an additional planet in these systems could transit.

\section{Acknowledgments} We are extremely grateful to the {\it EPOXI}  Flight and Spacecraft Teams that made these difficult observations possible.  At the Jet Propulsion Laboratory, the Flight Team has included M. Abrahamson, B. Abu-Ata, A.-R. Behrozi, S. Bhaskaran, W. Blume, M. Carmichael, S. Collins, J. Diehl, T. Duxbury, K. Ellers, J. Fleener, K. Fong, A. Hewitt, D. Isla, J. Jai, B. Kennedy, K. Klassen, G. LaBorde, T. Larson, Y. Lee, T. Lungu, N. Mainland, E. Martinez, L. Montanez, P. Morgan, R. Mukai, A. Nakata, J. Neelon, W. Owen, J. Pinner, G. Razo Jr., R. Rieber, K. Rockwell, A. Romero, B. Semenov, R. Sharrow, B. Smith, R. Smith, L. Su, P. Tay, J. Taylor, R. Torres, B. Toyoshima, H. Uffelman, G. Vernon, T. Wahl, V. Wang, S. Waydo, R. Wing, S. Wissler, G. Yang, K. Yetter, and S. Zadourian.  At Ball Aerospace, the Spacecraft Systems Team has included L. Andreozzi, T. Bank, T. Golden, H. Hallowell, M. Huisjen, R. Lapthorne, T. Quigley, T. Ryan, C. Schira, E. Sholes, J. Valdez, and A. Walsh.

Support for this work was provided by the {\it EPOXI} Project of the National Aeronautics and Space Administration's Discovery Program via funding to the Goddard Space Flight Center, and to Harvard University via Co-operative Agreement NNX08AB64A, and to the Smithsonian Astrophysical Observatory via
Co-operative Agreement NNX08AD05A.


\end{document}